\documentstyle[aps,prb,multicol,graphicx]{revtex}

\begin{document}
\draft
\title{Thermal conductivity of
pure and Mg-doped CuGeO$_3$
in the incommensurate phase
}
\author{J. Takeya, I. Tsukada and Yoichi Ando}
\address{Central Research Institute of Electric Power
Industry, Komae, Tokyo 201-8511, Japan}
\author{T. Masuda
and K. Uchinokura
}
\address{Department of
Advanced Materials Science,
The University of Tokyo,
Bunkyo-ku, Tokyo 113-8656, Japan}

\author{I. Tanaka \cite{yamanashi} and R. S. Feigelson}
\address{Center for Material Research, Stanford University,
Stanford, CA 94305}

\author{A. Kapitulnik}
\address{E. L. Ginzton Laboratory, Stanford University,
Stanford, CA 94305}

\date{Received \today}

\maketitle

\begin{abstract}

The thermal conductivity $\kappa$ of Cu$_{1-x}$Mg$_x$GeO$_3$ is
measured in magnetic fields up to 16~T.
At the transition
field $H_c$ to the high-field incommensurate (I) phase, $\kappa$
abruptly decreases.
While $\kappa$ of the pure CuGeO$_3$ is enhanced
with the application of higher magnetic fields,
an anomalous plateau feature shows up in the $\kappa(H)$
profile in the I phase of the Mg-doped samples which includes
antiferromagnetic ordering (I-AF phase).
With the help of specific heat data,
taken supplementally for the identical samples,
the above features are well understood
that phonons are strongly scattered by spin solitons
and that $\kappa$ in the I phase is governed by their
spatial distribution.
In particular, the plateau feature in the doped samples
suggests \lq\lq freezing" of the solitons,
in which antiferromagnetism must be playing an essential role
because this feature is absent at temperatures above $T_N$.

\end{abstract}
\pacs{PACS numbers: 66.70.+f, 75.30.Kz, 75.50.Ee}

%

\begin{multicols}{2}

\section{Introduction}
\label{sec:intro}

Recently, a new class of commensurate-incommensurate (C-I)
transition, magnetic-field induced C-I transition,
which had been theoretically predicted,\cite{cross_etc}
was experimentally established in an one-dimensional
antiferromagnetic (AF) Heisenberg spin system coupled to lattice
degrees of freedom, the spin-Peierls (SP) system;
the magnetic transition is observed in magnetization measurements
\cite{bloch_etc,hase0}
and the incommensurate lattice modulation is visibly shown in
x-ray diffraction experiments.\cite{kiryukhin0,kiryukhin}
The SP system shows a transition at $T_{\rm SP}$ from
the uniform (U) spin-1/2 quantum liquid phase
to the dimerized (D) phase,
where spin-singlet ground state is formed
at the expense of the lattice deformation
energy.\cite{bray}
With the application of magnetic field,
a new magnetic phase appears above a
threshold field $H_c$
because the nonmagnetic spin-singlet state becomes
energetically unfavorable in
high magnetic fields.
The magnetic moment is partly recovered in the chain,
and \lq\lq spin solitons" are formed in this magnetic phase.
The \lq\lq spin soliton" contains
both lattice deformation and moment (carrying spin 1/2),
whose spatial distribution is governed by the sine-Gordon equation.
Usually, the periodicity of the solitons are
determined simply by the applied magnetic field, and
as a result,
the spin and lattice modulation becomes
incommensurate.
Therefore, this high-field phase is called incommensurate (I) phase.
After the existence of the incommensurate modulation
is established,
the subsequent interest on this I phase is
how the spin and lattice modulation evolves
under higher magnetic fields (well above $H_c$),
where the soliton distance
becomes comparable to the soliton width.
\cite{horvatic,lorenz3,uhrig}

The discovery of the inorganic SP compound CuGeO$_3$
(Ref.~\onlinecite{hase1})
opened the way to impurity substitiution
and it turned out that
just a small amount of the impurity modifies
the phase diagram to a more complicated one;
\cite{hase2,manabe}
with decreasing temperature,
after the appearance of D phase at $T_{\rm SP}$,
the system shows another transition
to a peculiar ordered phase,
where
antiferromagnetic (AF) long-range order
and the SP order coexist
[dimerized AF (D-AF) phase].\cite{masuda}
The staggered moment and the lattice deformation
are spatially modulated in D-AF phase;
the amplitude is larger near the impurity sites
and smaller in between them.\cite{fukuyama,kojima}
In contrast to such well-established pictures of
the ordered phase in low fields (D and D-AF phase),
the spin distribution in high-field phases
of the impurity-doped systems
is still under question.
Also, the phase boundaries for the high-field
phases are not yet investigated for the system with its Cu site
substituted by impurities (Cu-O chains are directly disordered),
although there exist several reports on
these phase boundaries for the system
whose Ge site is substituted.
Recently, we reported that $\kappa$ of
Cu$_{1-x}$Mg$_x$GeO$_3$ is sensitive
to the Mg-doping (disorder),\cite{takeya,takeya2}
and demonstrated that such transport measurement is
useful in examining the disorder effects.
In this work,
we have measured the thermal conductivity of
Cu$_{1-x}$Mg$_x$GeO$_3$ single crystals
with the application of magnetic field up to 16 T.
The specific heat is also measured supplementally
for the identical samples.
First, $\kappa$ of the pure CuGeO$_3$ is measured
in order to identify the dominant scattering mechanism
in the I phase.
It turned out that
the dominant heat carriers (phonons)
are strongly scattered with the appearance
of the spin solitons in the I phase,
indicating that $\kappa$ is a good probe of the soliton
distribution.
Our main result is an anomalous plateau feature found in
the $\kappa(H)$ profile in the antiferromagnetically ordered
I (I-AF) phase of the Mg-doped samples,
suggesting that the spin solitons are frozen in the I-AF phase.

\section{Experimental}

The single crystals of Cu$_{1-x}$Mg$_x$GeO$_3$
were grown with a floating-zone method.
We have prepared both pure CuGeO$_3$
(the SP transition temperature $T_{\rm SP}$=14.5 K)
and two Mg-doped samples $x = 0.016$ and $x = 0.0216$
with $T_{\rm SP}$'s of 11.5~K and 9~K, respectively.
(The $T_{\rm SP}$'s are determined by
dc magnetization measurements).
The Mg concentration is carefully determined by
inductively coupled
plasma-atomic emission spectroscopy (ICP-AES). \cite{masuda2}
The Mg-doped samples shows the N\'{e}el transition
at $T_N =$~2.5~K and 3~K, respectively.\cite{masuda,masuda2}
Masuda {\it et al.} carefully examined $x$ dependence of
the $T_N$ for the same series of crystals
and found that $T_N$ jumps at the impurity-driven
transition from the D-AF phase to the uniform AF phase
without dimerization.
\cite{masuda,masuda2}
Note that the D-AF phase shows up below $T_N$ in the sample
used for our present study
because the Mg concentration of the both samples
is smaller than the critical
concentration $x_c$ ($= 0.023 \sim 0.027$).\cite{masuda,masuda2}

The thermal conductivity is measured using a
\lq\lq one heater, two thermometers" method.
The direction of the
heat current and the magnetic field is
parallel to the $c$-axis, i.e., the chain direction.
The detailed experimental technique is described
elsewhere.\cite{takeya,ando}
Since the total thermal conductivity is expressed as
$\kappa = \sum C_i D_i$,
where $C_i$ and $D_i$ are
specific heat and diffusivity of each heat carrier denoted by
$i$
(the diffusivity $D_i$ is proportional to the mean free path
of the heat carrier),
the specific heat measurement is helpful
in the analysis of the $\kappa$ data.
We have measured the specific heat in magetic fields
up to 16 T by the usual relaxation method
for the same samples.
The addenda consists of a sapphire substrate,
a cernox temperature sensor (calibrated in magnetic fields) and
1-k$\Omega$ micro heater.
The connection from the addenda to the heat sink
is made by Sb-Au thin wires to provide a moderate heat-relaxation
rate,
from which the heat capacity is calculated.

\section{Results}

\subsection{Pure CuGeO$_3$}
\label{subsec:pure}

\begin{figure}
\includegraphics[width=0.8 \columnwidth]{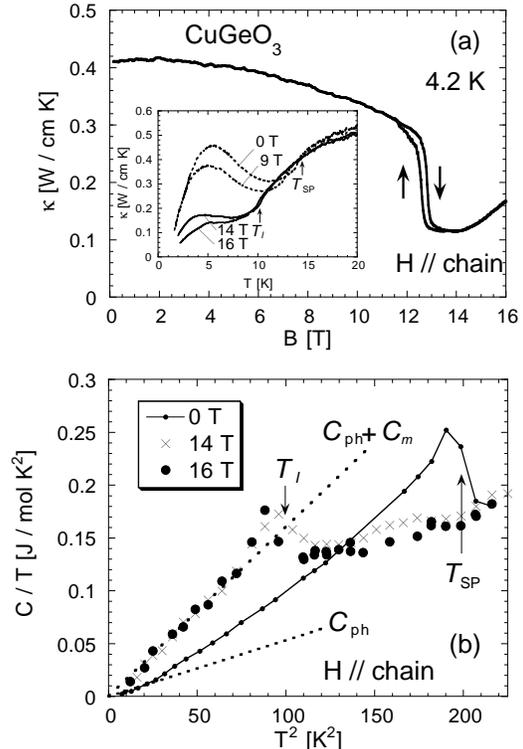}
\caption{(a) Magnetic field dependence of the
thermal conductivity of CuGeO$_3$ single crystal at 4.2 K.
The magnetic field is applied in the $c$-axis direction.
First order transition from the spin-Peierls to
the incommensurate phase takes place at $H_c$.
Inset: temperature dependence of the
thermal conductivity of CuGeO$_3$ in magnetic fields.
$T_{\rm SP}$ and $T_I$ are the transition temperatures
below which spin-Peierls phase and incommensurate phase
appear, respectively.
(b) $C / T$ in magnetic fields is plotted as a function of $T^2$.
The dotted line for the 0-T curve indicates the $T^3$ dependence of
the phonon contribution $C_{\rm ph}$.
The other dotted line for the $C / T$ curves in the I phase shows that
the additional magnetic contribution $C_m$ also
depends on temperature as $T^3$.
}
\label{fig1}
\end{figure}

The inset of Fig. 1(a) shows the temperature
dependence of $\kappa$ in fields up to 16 T.
The transition temperatures from the U phase
to the D phase (below 9~T)
and to the I phase (above 14 T) are indicated as
$T_{\rm SP}$ and $T_I$, respectively.
While both phonons and spin
excitations carry a large amount of heat in the U phase,
\cite{takeya,ando}
phonons become the dominant heat carrier
at temperatures well below $T_{\rm SP}$ in the D phase
because of the spin gap.
These phonons are mostly scattered by defects at the lowest
temperatures,
where spin excitations are almost absent, and
field dependence is small there;
the 0-T and 9-T curves merge below $\sim$~3~K.
When thermally excited spin excitations increase
with temperature and the scattering rate of the phonons are enhanced,
via spin-phonon coupling, $\kappa$ starts to decrease with temperature,
producing a pronounced peak around $\sim$~5~K.\cite{ando}
Since
this peak in $\kappa$ is a manifestation
of the spin gap,
the peak height is rapidly suppressed with magnetic
fields.\cite{takeya}

The field dependence of $\kappa$ at 4.2 K is presented
in the main panel of Fig. 1(a)
to see the suppression of the peak under magnetic fields.
A sudden drop in $\kappa$, accompanied by hysteresis, shows up
at the first order transition from the D phase to
the I phase ($H_c \sim 12.5$ T).
When the magnetic field increases above $H_c$,
$\kappa$ starts to increase at $\sim$ 14 T.
The specific heat $C$ is measured
for the same sample
in order to elucidate the origin of the field dependence of $\kappa$
($\kappa_i = C_i D_i$).
The 14-T and 16-T data are
plotted together with zero-field data in Fig. 1(b).
In contrast to $\kappa$,
$C$ in the I phase is {\it larger} than
that in the D phase below $T_I \sim$~10~K,
indicating that the jump in $\kappa$ at $H_c$
is caused by a sudden change in {\it diffusivity}.
It is also to be noted that the values of $C$ at 14~T
and at 16~T
do not significantly differ  below $T_I$,
while $\kappa$ shows an apparent increase above 14~T,
suggesting the enhancement in the diffusivity $D_i$
with the field application above $\sim 14$~T.

\subsection{Mg-doped CuGeO$_3$}

\begin{figure}
\includegraphics[width=0.8 \columnwidth]{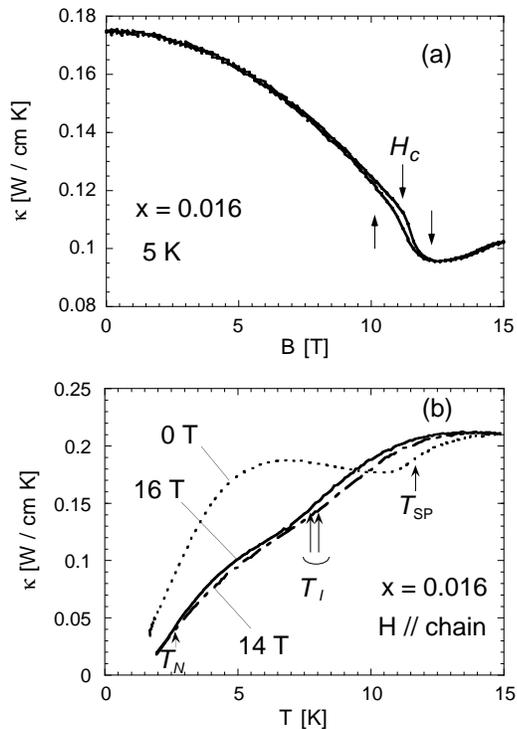}
\caption{(a) Magnetic field (along the CuO chain) dependence of the
thermal conductivity of Cu$_{1-x}$Mg$_x$GeO$_3$ with $x = 0.016$
at 5~K.
(b) Temperature dependence of $\kappa$ in magnetic fields up to 16~T
for the $x = 0.016$ sample.
$T_N$ indicates the N\'{e}el transition in the I phase.}
\label{fig2}
\end{figure}

Figure 2(a) shows the $\kappa(H)$ profile
of the $x = 0.016$ sample at 5 K,
which is above the zero-field $T_N$ ($\sim 2$~K),
determined from dc magnetization.
Similarly to the pure sample [Fig. 1(a)],
$\kappa$ is rapidly suppressed
with magnetic field when the field approaches
$H_c \sim 11.5$~T, and then
increases in the field range above $H_c$.
Hysteresis is observed in the vicinity of $H_c$
also for this sample,
although the transition becomes broader.

The temperature dependence of $\kappa$ is shown in Fig. 2(b).
The transition temperatures $T_{\rm SP}$ and $T_I$
are indicated with arrows (together with $T_N$).
Note that the 14-T and 16-T curves merge below $T_N$,
suggesting field independent $\kappa(H)$ above $H_c$
in this temperature range.
Since this feature is expected to be more pronounced
for a sample with higher $T_N$,
we measured the field dependence of the $x = 0.0216$ sample.
Figure 3(a) presents $\kappa(H)$ profiles
at the temperatures of 2.5~K and 5~K.
An apparent
plateau is observed above $H_c \sim 11.5$~T in the 2.5-K curve,
while the 5-K curve shows a slight upturn above $H_c$,
similarly to the field profile in the I phase of the pure sample.
Figure 3(b) shows the temperature dependence of $\kappa$
in three different fields above $H_c$
and that without magnetic field.
One can see that the 12-, 14- and 16-T curves merge
below $\sim 4$~K,
which is close to zero-field $T_N$ ($\sim$~3~K).

\begin{figure}
\includegraphics[width=0.8 \columnwidth]{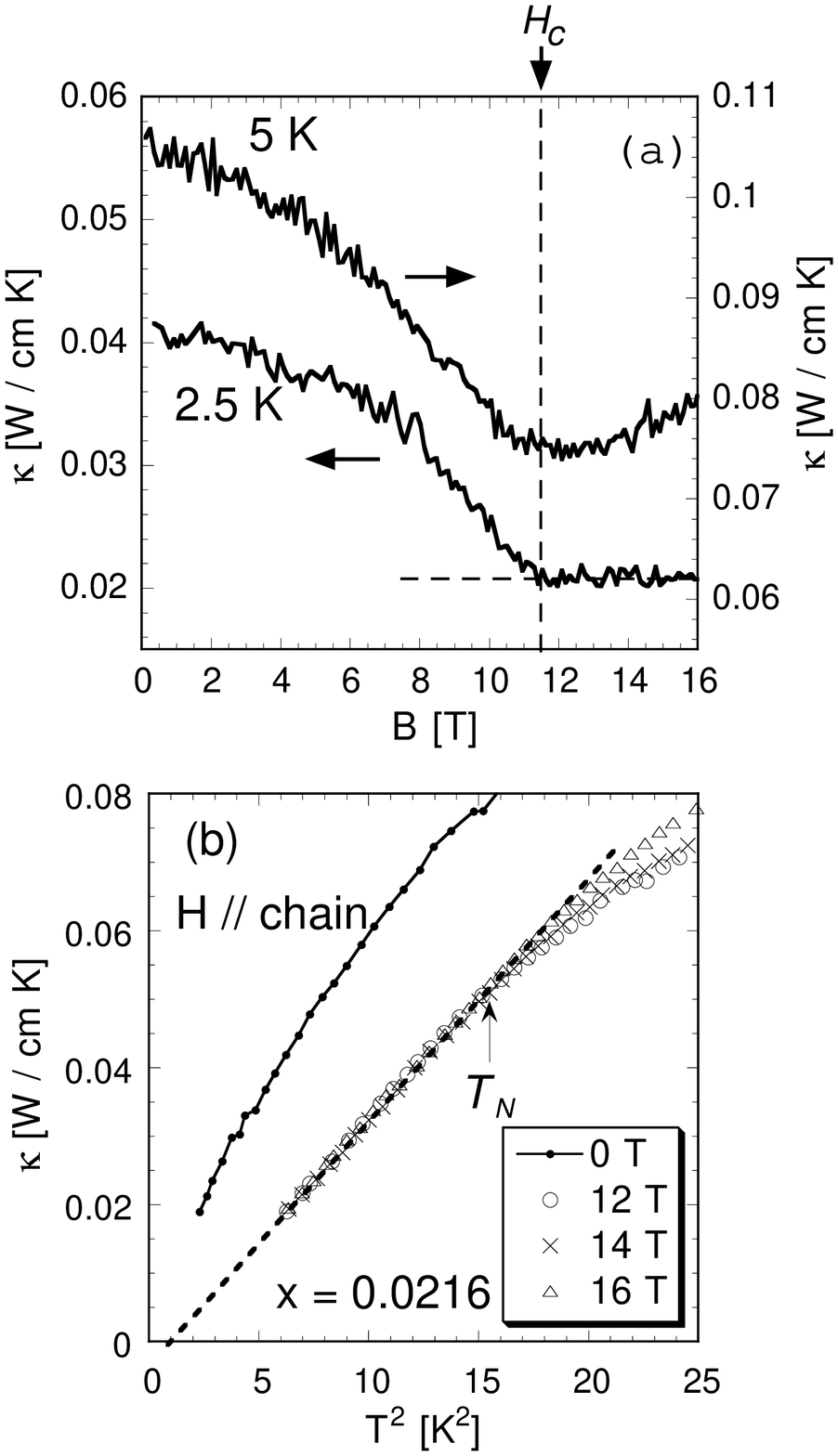}
\caption{(a) Magnetic field dependence of the
thermal conductivity of Cu$_{1-x}$Mg$_x$GeO$_3$ with $x = 0.0216$
at 2.5 K. $\kappa(H)$ becomes independent of magnetic fields in the
I phase (above $H_c$).
(b) $\kappa$ in magnetic fields up to 16~T is plotted against $T^2$
for the $x = 0.0216$ sample.
Below $T_N$, all the curves for the I phase merges to a
single line.
The dotted line shows that $\kappa$ is mostly proportional to
$T^2$ below $T_N$.}
\label{fig3}
\end{figure}

Figure 4 shows the
temperature dependence of the specific heat
of the same $x = 0.0216$ sample.
The 16-T curve shows
a $\lambda$-shaped anomaly,
indicating that $T_N$ under this magnetic field is around 4~K,
just below which the plateau feature appears in the $\kappa(H)$
profile.
Note that $T_N$ at 16~T is apparently higher than the zero-field
$T_N$, because the spin fluctuation
is reduced by field application.\cite{masuda3,hiroi}
It is also notable that a broad anomaly is observed
above $T_N$ up to $\sim $~8.5~K,
suggesting the existence of another phase at higher temperatures.

\begin{figure}
\includegraphics[width=0.8 \columnwidth]{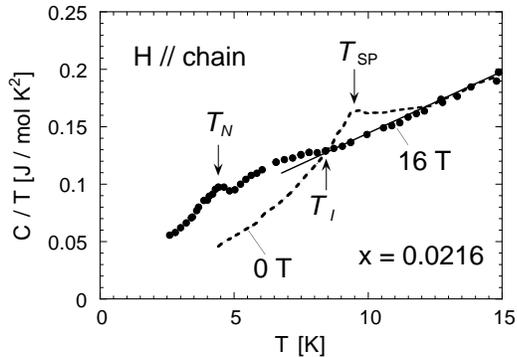}
\caption{Specific heat of the $x = 0.0216$ sample.
The I phase and the I-AF phase appear below $T_I$
and $T_N$, respectively.
The solid line is a guide to determine $T_I$.}
\label{fig4}
\end{figure}

\section{Discussion}

\subsection{Soliton scattering in I phase}

First, let us consider the $\kappa(H)$
profile at 4.2~K shown in Fig. 1(a).
The dominant heat carriers
are phonons in the pure CuGeO$_3$ well below $T_{\rm SP}$,
where the spin excitations are hardly excited
because of the spin gap.
Therefore, $\kappa(H)$ shown in Fig. 1(a)
mostly reflects the field dependence of $\kappa_{\rm ph}$
at least below $H_c$.
When the population of the spin excitations
gradually increases with field,
the scattering by the spin excitations increases and,
as the result,
$\kappa_{\rm ph}$ slowly decreases.

While the major part of the specific heat is due to phonons
in the D phase,
the additional component,
i.e., magnetic specific heat $C_m$,
appears in the I phase, as shown in Fig. 1(b).
This result reproduces the report by Lorenz {\it et al.},
\cite{lorenz2}
and is consistent with the theory claiming that $C_m$ is due to
magnetoelastic excitations called phasons.
\cite{bhattacharjee}
Since both $C_m$
and $C_{\rm ph}$
depends on temperature as $\propto T^3$,
the total $C$ is $\propto T^3$.
The thermal conductivity is
given by $\kappa = \sum C_i D_i$ ($i = {\rm ph},\ m$),
where
the additional magnetic specific heat $C_m$ reaches twice as large as
the phonon contribution $C_{\rm ph}$ [see Fig.~1(b)].
Nevertheless,
$\kappa$ suddenly {\it decreases} at the transition
from the D phase to the I phase, meaning that
the decrease in $\kappa_{\rm ph}$ is dominant at the transition,
though the newly appearing phasons could also carry some heat.
Since little change in $C_{\rm ph}$ is expected at $H_c$,
it must be $D_{\rm ph}$ that is responsible for the
abrupt change in $\kappa(H)$,
indicating the sudden appearance of new scatterers of phonons
in the I phase.
It is most likely that phonons are strongly scattered by
domain walls accompanied with the spin solitons,
which would be equivalent to the scattering by the phasons.
Note that the ultra-sonic results,
in which sudden change in elastic constants was shown at $H_c$,
are understood in terms of the phonon-soliton interaction.
\cite{saint_paul}

The $\kappa(H)$ profile above $H_c$ also is not simply
explained by $C$ in the I phase;
$\kappa$ increases above $\sim$~14~T,
while the difference in $C_m$ at 14 T and that at 16 T
is negligible within
the accuracy of our measurement ($\sim 10$\% of $C_m$),
\cite{remenyi} as shown in Fig. 1(b).
Since the enhancement in $\kappa$ from 14 T to 16 T
is more than four times larger than this maximum error,
the diffusivity should be responsible
for the magnetic-field dependence of $\kappa$ also in the
field range above $H_c$.

Assuming that $D_{\rm ph}(H)$ is governed by the scattering by the
domain walls in the I phase, as is discussed above,
$D_{\rm ph}(H)$ should be modified when the soliton distribution
changes with field.
In a sufficiently high magnetic field
where the soliton distance $d$ becomes comparable to
the soliton width $\xi$,
the adjacent solitons would overlap with each other and
the lattice and spin modulation might become sinusoidal wave,
as is suggested by a recent NMR measurement \cite{horvatic,uhrig}
and a magnetostriction result.\cite{lorenz3}
It is expected that the cross section of the domain-wall scattering
is smaller for the sinusoidal-wave modulation than
for the independent solitons,
because the amplitude of the lattice modulation is smaller for
the sinusoidal-wave distribution.
Such field dependence of the soliton distribution
well explains the enhancement in $D_{\rm ph}(H)$ with field above $H_c$,
and is consistent with the $\kappa(H)$ profile in the I phase.
The ratio of $d/\xi$ is $\sim 2.3$ at $\sim 14$~T
(Ref.~\onlinecite{horvatic})
where the overlapping becomes evident in $\kappa(H)$.

\subsection{Incommensurate phases in the Mg-doped samples}

Since $\kappa$ and $C$ data demonstrate anomalies
at the transition temperatures and at the critical field,
we can discuss the phase boundary for the high-field phases
of the Mg-doped CuGeO$_3$.
So far, contradictory results are reported
on the high-field incommensurate phases
for CuGeO$_3$ crystals whose Ge site is substituted by Si.
Two transitions are observed in the specific heat measurement;
one is from the U phase to the incommensurate phase without AF ordering
(I phase)
and the other is a N\'{e}el transition from this I phase to the
I-AF phase.\cite{hiroi}
On the other hand, the result of a ultrasound-velocity measurement
suggests a direct transition from the U phase to the I-AF phase.
\cite{poirier}

Let us discuss this problem on the basis of our $\kappa$ and $C$ data
for the Mg-doped samples,
where Cu-O chain is directly modulated.
The N\'{e}el transition to the I-AF phase
is apparent also for the Mg-doped samples;
the specific heat shows a clear $\lambda$-shaped peak (Fig. 4),
and the field dependence of $\kappa$ disappears
as shown in Fig.~2(b) and Fig.~3(b).
In addition to the features at the N\'{e}el ordering,
an enhancement is observed below $T_I \sim 9$~K in $\kappa$
for the $x = 0.016$ sample [Fig. 2(b)].
The specific heat data for the $x = 0.0216$ sample in 16~T
also shows an enhancement below $T_I \sim 8.5$~K (Fig. 4).
The results indicate the existence of the I phase without
AF order for the Mg-doped samples.\cite{masuda3com}
However, the anomaly in the $C$ data is not so obvious at $T_I$
as at $T_N$.
This is probably because of the existence of
short-range ordered (SRO) state
in the SP transition in the impurity-doped CuGeO$_3$.
Noting that long-range SP order is not established
down to $\sim$~6.5~K at 0~T for the $x = 0.0216$ sample
(Ref.~\onlinecite{masuda2,wang}),
and assuming the SRO state also in magnetic fields above $H_c$,
the \lq\lq transition temperature $T_I$" will not be well-defined.

\begin{figure}
\includegraphics[width=0.8 \columnwidth]{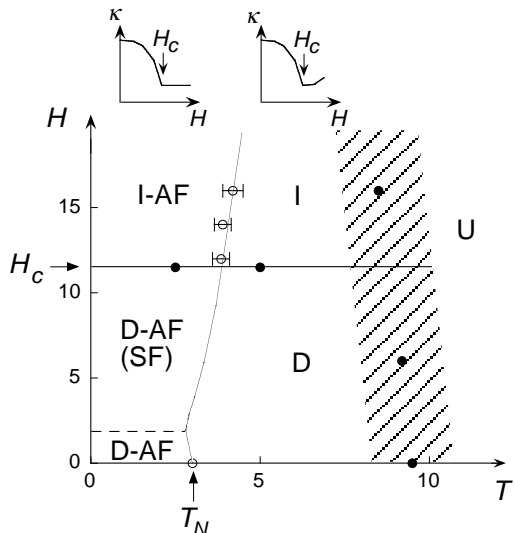}
\caption{The schematic $H$-$T$ phase diagram
of the $x = 0.0216$ sample,
based on the results of the $\kappa$ and $C$ measurements.
Open and closed circles indicate the transition temperatures
determined from the $\kappa$ and $C$ data.
The hatched area indicates the short-range ordered regime.}
\label{fig5}
\end{figure}

The field induced transition
from D to I and from D-AF to I-AF
phases at $H_c$ is also evident in the $\kappa(H)$ profile;
$\kappa$ for the $x = 0.016$ sample and
the two curves for the $x = 0.0216$ sample
(at 2.5 K and 5 K) rapidly decrease
(with hysteresis when $x = 0.016$)
with the application of magnetic fields
below $H_c \sim$~11.5~T,
while the decrease abruptly disappears above $H_c$
[Fig. 2(a) and Fig. 3(a)],
similarly to the D-I transition observed for the pure CuGeO$_3$.
Note that the transition at 5~K (above $T_N \sim$~4~K)
and that at 2.5~K (below $T_N$) take place
at almost the same field around $H_c \sim$~11.5~T, i.e.,
the value of $H_c$ is irrelevant to
whether the temperature is above or below $T_N$.
The above results suggest that both transitions,
from D to I and from D-AF to I-AF,
are \lq\lq commensurate-incommensurate" transition,
which
is characterized by the appearance of the spin solitons.
The suggested phase diagram is presented in Fig. 5,
following the discussion in this subsection.

\subsection{Freezing of the solitons in I-AF phase}

The most striking observation in the $\kappa(H)$ profile for
the Mg-doped samples is the plateau feature which appears in the I-AF
phase [see Fig. 3(a)].
Since the plateau suddenly appears above $H_c$,
it should be related to the specific scattering mechanism
in the incommensurate phase, i.e.,
the scattering by the domain-walls.
Therefore,
the field independence in the I-AF
phase suggests that the soliton population and
their distribution are {\it not}
affected by the magnitude of the applied magnetic field,
indicating that the solitons are frozen.
Since the plateau in $\kappa(H)$ is present only
below $T_N$ [see Fig. 3(b)],
it is the specific feature of the I-AF phase, and
the AF order must be essential for the
soliton freezing.

In the I-AF phase,
it is suggested that solitons are preferably
distributed around the impurity sites and that
staggered moments are distributed also
in between these solitons.\cite{saito}
Since the spin-flop transition takes place at $\sim 1$~T,
\cite{hase4,nojiri}
the staggered moments are directed almost
perpendicular to $H$ in the I-AF phase.
When the magnetic field increases,
there are two ways to gain the magnetic
energy $\Delta M H$, where $\Delta M$ is the additional
magnetization:
one is to append other solitons in between the impurity sites,
though this position is not so favored compared to the impurity
sites, and the other is
to cant the staggered moments to the direction of $H$.
The second process
can give an origin of the soliton \lq\lq freezing"
induced by the antiferromagnetism,
because the soliton distribution
is preserved under further magnetic field.
Although the above scenario gives a plausible explanation,
independent measurements by neutron scattering and/or
NMR are
required to elucidate this microscopic
mechanism of the antiferromagnetism-induced soliton freezing.

The dotted line in Fig.~3(b) indicates that
$\kappa$ is proportional to $T^2$ when the sample is cooled below $T_N$.
Since the specific heat have
the temperature dependence of $\propto T^3$,
the diffusivity should behave as $\propto 1/T$,
the typical temperature dependence of $\kappa_{\rm ph}$ when phonons
are scattered by dislocations of solids, like defects.
The observation shows that the domain walls
of the spin solitons behave as rigid dislocations,
which would be consistent with the above picture of the
frozen solitons in the I-AF phase.

\section{Summary}

The results of the thermal conductivity measurement on
both pure and Mg-doped CuGeO$_3$,
accompanied with the specific heat measurement for the same crystals,
help to further understand
the high-field incommensurate phase of the SP system,
the spin-soliton phase.
First, it turned out that the phonon transport in the I phase is
governed by the domain-wall scattering
due to the significant phonon-soliton interaction, and
is strongly influenced by spatial distribution of the solitons.
Since the kink features in $H$ and $T$ dependence
caused by the appearance (change in the distribution)
of the solitons tell us the transition
into (in) the I phase,
the $H - T$ phase diagram of the Mg-doped CuGeO$_3$
(where the CuO chain is directly disturbed), is established.
Based on our result of $\kappa(H)$ above $H_c$,
it is suggested that
the spin solitons are frozen in the I-AF phase
with the help of the AF ordering,
providing a clue for the full understanding
of the spin distribution in this unusual coexistence phase.

\section{Acknowledgment}

We greatly thank M. Saito for fruitful discussions.

%
%

%

\end{multicols}

\end{document}